\documentclass[oneoclumn,preprintnumbers,prd,preprint,superscriptaddress,nofootinbib,12pt,a4paper]{revtex4-1}

\usepackage{graphicx}
\usepackage{amsmath,amsfonts,amssymb}
\usepackage{epstopdf}
\usepackage{natbib}
\usepackage{hyperref}
\usepackage[margin=1in,footskip=0.2in]{geometry}
\usepackage{color,verbatim}
\hypersetup{colorlinks   = true,
            urlcolor     = blue,
            citecolor    = blue,
            linkcolor    = blue,
            menucolor    = blue,
            anchorcolor  = blue,
            filecolor    = blue}
\enlargethispage{\baselineskip}

\date{\today}

\usepackage[normalem]{ulem}
            
\begin{document}

\title{Note on fundamental physics tests from black hole imaging: Comment on ``Hunting for extra dimensions in the shadow of Sagittarius A$^*$''}
\date{\today}

\author{Sunny Vagnozzi}
\email{sunny.vagnozzi@ast.cam.ac.uk}
\affiliation{Kavli Institute for Cosmology (KICC), University of Cambridge,\\ Madingley Road, Cambridge CB3 0HA, United Kingdom}

\author{Luca Visinelli}
\email{luca.visinelli@sjtu.edu.cn}
\affiliation{Tsung-Dao Lee Institute (TDLI), 520 Shengrong Road, 201210 Shanghai, P.\ R.\ China}
\affiliation{School of Physics and Astronomy, Shanghai Jiao Tong University, 800 Dongchuan Road, 200240 Shanghai, P.\ R.\ China}

\begin{abstract}
\noindent Several works over the past years have discussed the possibility of testing fundamental physics using Very Long Baseline Interferometry horizon-scale black hole (BH) images, such as the Event Horizon Telescope (EHT) images of M87$^*$ and Sagittarius A$^*$ (Sgr A$^*$), using the size $r_{\rm sh}$ and deviation from circularity $\Delta \mathcal{C}$ of the BH shadow. For the case of the EHT image of Sgr A$^*$, limits on $\Delta \mathcal{C}$ are not available due to the sparse interferometric coverage of the 2017 observations, alongside the short variability timescale of Sgr A$^*$ compared to M87$^*$. Concerning this point, we comment on the results of a recent preprint which purports to have derived new limits on extra dimensions using the deviation from circularity of Sgr A$^*$'s shadow. The latter is quoted to be $\lesssim 10\%$ as with M87$^*$, based on the ``similarity'' of the two shadows: however, this is an incorrect assumption, invalidating the subsequent results. In the immediate future, the simplest tests of fundamental physics from Sgr A$^*$'s image will therefore mostly have to rely on $r_{\rm sh}$, whereas additional observables such as the photon ring and azimuthal angle lapse should soon be available and allow for novel tests.
\end{abstract}

\maketitle

Millimeter-scale Very Long Baseline Interferometry (VLBI) has recently provided us with the first horizon-scale images of supermassive black holes (BHs): these feature a central brightness depression, related to the underlying BH shadow~\cite{Luminet:1979nyg,Falcke:1999pj}, surrounded by a bright ring of emission. In most scenarios, the BH shadow corresponds to the apparent image of the photon region, the region of space-time where photons are forced to travel along orbits~\cite{Cunha:2018acu,Perlick:2021aok,Wang:2022kvg}. Modifications to General Relativity, the presence of additional fields, as well as violations of fundamental principles, typically leave imprints on the BH shadow: horizon-scale BH images can thus be used to test fundamental physics, provided the size of the bright ring can be used as a proxy for the shadow size $r_{\rm sh}$, with little dependence on the details of the accretion flow, which is the case for the geometrically thick, optically thin, radiatively inefficient accretion flows surrounding both M87$^*$ and Sagittarius A$^*$ (Sgr A$^*$)~\cite{Narayan:2019imo,Volkel:2020xlc,Bronzwaer:2021lzo,Lara:2021zth,Ozel:2021ayr,Younsi:2021dxe,Kocherlakota:2022jnz,EventHorizonTelescope:2022xqj}.

Over the past years, considerable work has indeed been devoted to tests of fundamental physics from the Event Horizon Telescope (EHT) horizon-scale images of M87$^*$~\cite{EventHorizonTelescope:2019dse,EventHorizonTelescope:2019ggy}, for a representative example see e.g.\ the EHT paper Ref.~\cite{EventHorizonTelescope:2021dqv}. For the case of M87$^*$, these tests have been based both on the size of the bright ring of emission (as a proxy for the BH shadow size $r_{\rm sh}$), as well as the deviation from circularity of the shadow $\Delta \mathcal{C}$ (e.g.\ Ref.~\cite{Bambi:2019tjh}): the latter quantifies the oblateness of the shadow, and is tied to deviations from the quadrupole moment of the Kerr metric~\cite{Johannsen:2010ru}. In Ref.~\cite{EventHorizonTelescope:2019dse}, the limit $\Delta \mathcal{C} \lesssim 10\%$ has been quoted, and this has been translated into a deviation from the Kerr quadrupole moment of order $4$.

In Ref.~\cite{Vagnozzi:2019apd}, we have shown how the previous limit can be used to test the properties of non-compactified, large extra dimensional scenarios such as the Randall-Sundrum (RS) model~\cite{Randall:1999ee,Randall:1999vf}. This is characterized by a five-dimensional anti-de Sitter space (AdS$_5$) brane-world with AdS$_5$ curvature radius $\ell$, where the extra dimension has an infinite size and a negative bulk cosmological constant. In the characterization proposed, gravity is free to propagate on the bulk of the AdS$_5$ spacetime, affecting the circularity of the BH shadow (see also Refs.~\cite{Banerjee:2019nnj,Neves:2020doc,Hou:2021okc}). We found that the limit $\Delta \mathcal{C} \lesssim 10\%$ translates to a limit of $\ell \lesssim 170\,{\rm AU}$, much weaker than constraints from precision gravity tests~\cite{Maartens:2010ar}.
 
Recently, Ref.~\cite{Wu:2022ydc} purported to have derived a new constraint on the AdS$_5$ curvature radius $\ell$ based on the new EHT horizon-scale images of Sgr A$^*$~\cite{EventHorizonTelescope:2022xnr}. This preprint, almost entirely based on our work in Ref.~\cite{Vagnozzi:2019apd} and following a similar analysis, quotes a deviation from circularity $\Delta \mathcal{C} \lesssim 10\%$ for Sgr A$^*$, analogously to M87$^*$. We disagree with the findings obtained in the preprint, as the procedure used to derive the results are not correct. The author quotes the limit $\Delta \mathcal{C} \lesssim 10\%$ on the basis of ``Sgr A$^*$ and M87$^*$ [being] very similar and [...] accurately described by the Kerr metric'' (see Footnote~1 of Ref.~\cite{Wu:2022ydc}). The problem is that limits on $\Delta \mathcal{C}$ have not been quoted for Sgr A$^*$'s image, as explicitly discussed by the EHT collaboration on page 19 of Ref.~\cite{EventHorizonTelescope:2022xqj}, due to the sparse interferometric coverage of the observations taken in 2017 when Sgr A$^*$ had been imaged. This is related to the status of the EHT network, which at the time of the observations did not have enough telescopes online to robustly obtain circularity measurements. Moreover, the variability timescales $t_g \sim GM/c^3$ change considerably between Sgr A$^*$ and M87$^*$ due to the sheer difference in mass, with the short timescale of Sgr A$^*$ leading to an extensive overnight variability. Therefore, it is not yet possible to perform a similar analysis as the one we have conducted in Ref.~\cite{Vagnozzi:2019apd} to obtain analogous results from Sgr A$^*$. In fact, some of the latest results based on the new image of Sgr A$^*$ do not make use of $\Delta \mathcal{C}$ (see e.g.\ Refs.~\cite{Chen:2022nbb,Jusufi:2022loj,Chen:2022lct,Vagnozzi:2022moj}).

To sum up, our past works with our collaborators has shown how horizon-scale images of supermassive BHs can be used to test fundamental physics, using the size $r_{\rm sh}$ and deviation from circularity $\Delta \mathcal{C}$ of the shadow. However, while for M87$^*$ information on both $r_{\rm sh}$ and $\Delta \mathcal{C}$ is available, for the 2017 EHT observations of Sgr A$^*$ only the former one is available: therefore, in the near future, the simplest tests of fundamental physics from Sgr A$^*$'s shadow will inevitably have to be based on its size and not its oblateness, until the EHT collaboration or its successor will explicitly constrain the latter. We have commented on an erroneous result obtained in Ref.~\cite{Wu:2022ydc} in the context of extra dimensions, based on incorrect premises relating to the deviation from circularity of Sgr A$^*$'s shadow, and taken the opportunity to draw the attention of the community to these related important points. In the future, the availability of more telescopes within VLBI networks will allow for even stronger tests of fundamental physics~\cite{Blackburn:2019bly,2021ApJS..253....5R}. Besides $r_{\rm sh}$ and $\Delta \mathcal{C}$, complementary observables and techniques to explore new physics are being developed, including polarimetric measurements, variations in the azimuthal angle lapse, and shadow drift~\cite{Chen:2019fsq,Hadar:2020fda,Chen:2022nbb}, all of which are already accessible, or potentially accessible soon, to current VLBI arrays.

\newpage
\begin{acknowledgments}
\noindent S.V.\ is supported by the Isaac Newton Trust and the Kavli Foundation through a Newton-Kavli Fellowship, and by a grant from the Foundation Blanceflor Boncompagni Ludovisi, n\'{e}e Bildt. S.V.\ acknowledges a College Research Associateship at Homerton College, University of Cambridge.
\end{acknowledgments}

\bibliography{note_on_circularity.bib}

\end{document}